\documentclass[a4paper,11pt]{article}

\usepackage{contribution}

\newcommand{\weblink}[2][]{%
    \ifthenelse{\equal{#1}{}}%
    {\textnormal{\url{#2}}}%
    {\textnormal{\href{#2}{#1}}}%
}

\newcommand{\acknowledgements}[1]{%
  \bigskip\bigskip
  \textsf{\textbf{\Large Acknowledgements}} \\[2ex]
  {#1}
  \bigskip
}

\def\beq{\begin{equation}}
\def\eeq#1{\label{#1}\end{equation}}
\def\eeqn{\end{equation}}

\def\beqa{\begin{eqnarray}}
\def\eeqa#1{\label{#1}\end{eqnarray}}
\def\eeqan{\end{eqnarray}}

\let\bar=\overbar

\def\Dslash{\not{\hbox{\kern-4pt $D$}}}
\def\dslash{\not{\hbox{\kern-2pt $\del$}}}

\def\msb{{\bar{\ssstyle M \kern -1pt S}}}

\newcommand{\contribution}[7][]{%
  \clearpage
  \thispagestyle{plain}
  \ifthenelse{\equal{#1}{}}
  {\hypersetup{pdftitle={#2}}}
  {\hypersetup{pdftitle={#1}}}
  \hypersetup{pdfauthor={{#3} {#4}}}
  {\centering\normalfont\LARGE\bfseries\sffamily #2 \par\nobreak}
  \lhead{}
  \chead{%
    \textit{\footnotesize XIV International Conference on Hadron Spectroscopy
      (\weblink[\textit{hadron2011}]{http://www.hadron2011.de}), 13-17 June 2011, Munich, Germany}%
  }
  \rhead{}
  \bigskip
  \begin{center}
    {#3} {#4}\ifthenelse{\equal{#6}{}}{}{\footnote{\weblink[#6]{mailto:#6}}}
    \ifthenelse{\equal{#7}{}}{}{#7} \\
    \textit{#5}
  \end{center}
  \bigskip
}

\renewcommand{\abstract}[1]{%
  \begin{center}
    \begin{minipage}{0.85\textwidth}
      \begin{footnotesize}
        #1
      \end{footnotesize}
    \end{minipage}
  \end{center}
  \bigskip
}

\begin{document}

%
%
%
%
%
{  


%

\contribution[Highlights from BESIII experiment]  
{Highlights from BESIII experiment}  
{Hai-Bo}{Li}  
{Institute of High Energy Physics \\
Bejing 100049, China}  
{lihb@ihep.ac.cn}  
{on behalf of the BESIII Collaboration}  
%

\abstract{%
BESIII had collected large data samples on $J/\psi$ and
$\psi^\prime$ peaks during the first run in 2009. From 2010 to 2011,
about 2.9 fb$^{-1}$ integrated luminosity were obtained on the peak
of $\psi(3770)$ for open charm physics. We review recent results on
charmonium decays and hadron spectroscopy. The prospects on open
charm physics are also discussed. }
%

\section{Introduction}

The newly built BEPCII/BESIII is an upgrade to the previous
BEPC/BES~\cite{besiii}. The BEPCII is a double ring collider with a
design luminosity of $1\times 10^{33}$cm$^{-2}$s$^{-1}$ at a
center-of-mass energy of 3.78 GeV, which luminosity is one order of
higher than that at CESR-c. It is operating between 2.0 and 4.6 GeV
in the center of mass. The BESIII experiment is used to study the
charm and $\tau$ physics. It is foreseen to collect on the order of
10 billion $J/\psi$ events or 3 billion $\psi(2S)$ events per year
according to the designed luminosity.  About 32 million
$D\overline{D}$ pairs and $2.0$ million $D_S \overline{D}_S$ at
threshold will be collected per year~\cite{besiii}.  In last run,
the peak luminosity of BEPCII has reached $6.4 \times
10^{32}$cm$^{-2}$s$^{-1}$.

The BESIII detector~\cite{besiii}
consists of the following main components:
1) a main draft chamber (MDC) equipped with about 6500 signal wires
and 23000 field wires arranged as small cells with 43 layers. The
designed single wire resolution is 130 $\mu m$ and the momentum
resolution 0.5\% at 1 GeV; 2) an electromagnetic calorimeter(EMC)
made of 6240 CsI(Tl) crystals. The designed energy resolution is
2.5\%@1.0 GeV and position resolution 6mm@1.0 GeV; 3) a particle
identification system using Time-Of-Flight counters made of 2.4 m
long plastic scintillators. The designed resolution is 80 ps for two
layers, corresponding to a K/$\pi$ separation (2$\sigma$ level) up
to 0.8 GeV; 4)a superconducting magnet with a field of 1 tesla; 5) a
muon chamber system made of Resistive Plate Chambers(RPC).

In 2009, the BESIII had collected about 225 M and 106M data set on
the J/$\psi$ and $\psi^\prime$ peaks, respectively. About 2.9
fb$^{-1}$ integrated luminosity on the $\psi(3770)$ peak had been
accumulated for open charm physics, and the data size is about 3.5
times of that at CLEO-c. The results in this paper is based above
data set.

\section{Highlights from BESIII}

The BESIII collaboration has published so-far a variety of papers
with many new results in the field of light hadron and charmonium
spectroscopy, as well as charmonium
decays~\cite{hc_paper,gP_paper,PP_paper,ppbar_paper,X1835_paper,matrix_paper,4pi_paper,
mixing_paper,gV_paper,vv_paper}. A number of new hadronic states
were discovered or confirmed, and various decay properties were
measured for the first time. In addition, many data analyses are in
an advanced stage and will lead to a rich set of new publications in
the near future. Here, we show recent highlights from the BESIII,
thereby, illustrating the potential of the BESIII experiment. In
this paper, for the reported experimental results, the first error
and second error will be statistical and systematic, respectively,
if they are not specified.

\subsection{$\eta_c(1S)$ resonance via $\psi^\prime \rightarrow \gamma \eta_c$ decay}

Precise measurement of M1 transition of $\psi^\prime$ is important
for us to understand the QCD in the relativistic and nonperturbative
regimes.  The $\psi^\prime \rightarrow \gamma \eta_c$ transition is
also a source of information on the $\eta_c$ mass and width. There
is currently a 3.3 $\sigma$ inconsistency in previous $\eta_c$ mass
measurements from J$/\psi$ and $\psi^\prime\rightarrow \gamma
\eta_c$ (averaging $2977.3 \pm 1.3$ MeV/$c^2$) compared to $\gamma
\gamma$ or $p\bar{p}$ production (averaging 2982.6 $\pm$1.0
MeV/$c^2$)\cite{ref:hc-hyper}. The width measurements also spread
from 15 to 30 MeV, it is around 10 MeV in the earlier days of
experiments using J/$\psi$ radiative transition~\cite{mark3,bes1},
while the recent experiments, including photon-photon fusion and B
decays, gave higher mass and much wider
width~\cite{cleo-c,babar,belle1,belle2}. The most recent study by
the CLEO-c experiment~\cite{cleo-c-new}, using both $\psi^\prime$
and J/$\psi \rightarrow \gamma \eta_c$ decays, and pointed out there
was a distortion of the $\eta_c$ line shape. CLEO-c attributed the
$\eta_c$ line-shape distortion to the energy-dependence of the M1
transition matrix element. In the $J/\psi \rightarrow \gamma \eta_c$
from CLEO-c, the distorted $\eta_c$ lineshape can be described by
the relativistic Breit-Wigner (BW) distribution modified by a factor
of E$^3_{\gamma}$ together with a dumping factor to suppress the
tail on the higher photon energies. KEDR Collaboration did the same
thing but tried different dumping factor~\cite{kedr}.

Based on the data sample of 106 M $\psi^\prime$ events collected
with BESIII detector,  the $\eta_c$ mass and width are measured from
the radiative transition $\psi^\prime \rightarrow \gamma \eta_c$.
The $\eta_c$ candidates are reconstructed from six exclusive decay
modes: $K_s K\pi$, $K^+K^-\pi^0$, $\eta \pi^+\pi^-$, $K_s
K^+\pi^-\pi^+\pi^-$, $K^+K^-\pi^+\pi^-\pi^0$, and 3$(\pi^+\pi^-)$,
where $K_s$ is reconstructed in  $\pi^+\pi^-$ mode, $\eta$ and
$\pi^0$ from $\gamma \gamma$ final states. For a hindered M1
transition the matrix element acquires terms proportional to
$E^2_{\gamma}$, which, when combined with the usual $E^3_{\gamma}$
term for the allowed transitions, lead to contributions in the
radiative width proportional to $E^7_{\gamma}$. Thus, the $\eta_c$
lineshape is described by a BW modified by $E_{\gamma}^7$ convoluted
with a resolution function. It is important to point out that the
interference between $\eta_c$ and non-resonance in the signal region
is also considered. The statistical significance of the interference
is 15 $\sigma$. This affects the $\eta_c$ resonant parameters
significantly. Assuming an universal relative phase between the two
amplitudes, we obtain $\eta_c$ mass and width, $M = 2984.2 \pm 0.6
\pm 0.5$ MeV/$c^2$ and $\Gamma = 31.4 \pm 1.2 \pm 0.6$ MeV,
respectively, as well as the relative phase $\phi = 2.41 \pm
0.06\pm0.04$ rad. Figure~\ref{fig:eta-c-shape} shows the fit results
in the six $\eta_c$ decay modes. With precise measurement of the
$\eta_c$ mass, one can obtain the hyperfine splitting, $\Delta
M_{hf}(1S)_{c\bar{c}} \equiv = M(J/\psi) - M(\eta_c) = 112.5 \pm
0.8$ MeV, which agrees with the quark model prediction~\cite{kseth},
and will be helpful for understanding the spin-dependent
interactions in hidden quarkonium states.

\begin{figure}[htb]
  \begin{center}
    \includegraphics[width=0.7\textwidth]{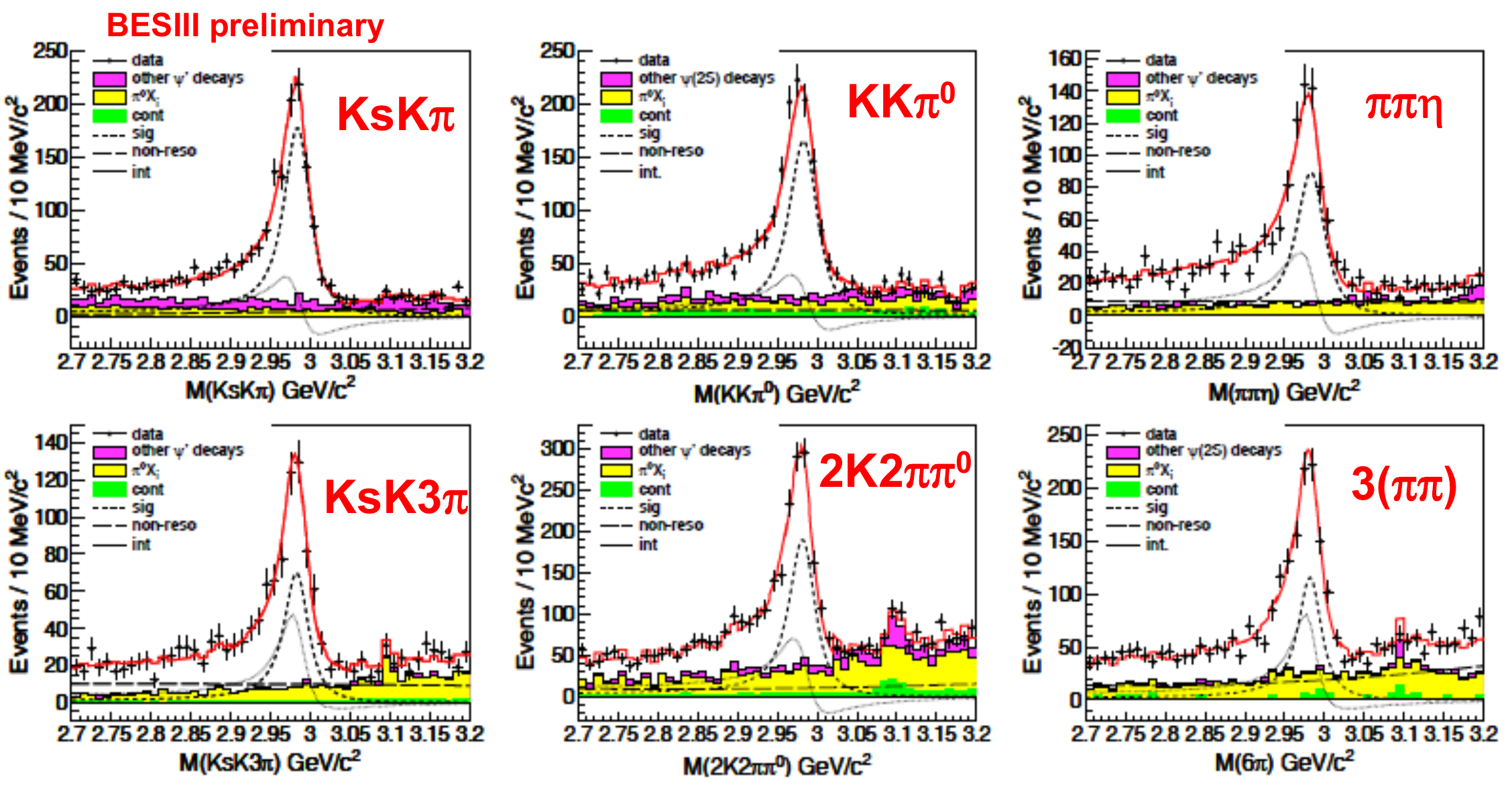}
    \caption{The invariant mass distributions for the decays $K_s K\pi$, $K^+K^-\pi^0$, $\eta \pi^+\pi^-$, $K_s
K^+\pi^-\pi^+\pi^-$, $K^+K^-\pi^+\pi^-\pi^0$, and 3$(\pi^+\pi^-)$,
respectively. Solid curves show the fitting results; the fitting
components ($\eta_c$ signal/non-resonance/interference) are shown as
(dashed/long-dashed/dotted) curves. Points with error bar are data,
shaded histograms are (in green/yellow/magenta) for (continuum/other
$\eta_c$ decays/other $\psi^\prime$ decays) backgrounds.}
    \label{fig:eta-c-shape}
  \end{center}
\end{figure}

\subsection{Observation of $\psi^\prime \rightarrow \gamma \eta_c(2S)$}

The first radially excited S-wave spin singlet state in the
charmonium system, $\eta_c(2S)$, was observed by the Belle
Collaboration in the decay process $B^\pm \rightarrow K^\pm
\eta_c(2S)$, $\eta_c(2S) \rightarrow K_s K^\pm
\pi^\mp$~\cite{belle-etac-2s}. It was confirmed by the
CLEO~\cite{cleo-etac-2s} and BABAR~\cite{babar-etac-2s}
Collaborations in the two-photon fusion process $e^+e^- \rightarrow
e^+e^- (\gamma\gamma), \gamma \gamma \rightarrow
\eta_c(2S)\rightarrow K_sK^\pm\pi^\mp$ and by the BABAR
Collaboration in the double-charmonium production process $e^+e^-
\rightarrow J/\psi (c\bar{c})$~\cite{babar-etac-2s-double}. The only
evidence for $\eta_c(2S)$ in the $\psi^\prime \rightarrow \gamma
\eta_c(2S)$ decay was from Crystal Ball
Collaboration~\cite{crystal-ball-etac-2s} by looking at the
radiative photon spectrum. Recently, CLEO-c Collaboration searched
for the $\psi^\prime \rightarrow \gamma \eta_c(2S)$ signal with
$\eta_c(2S)$ exclusive decay into 11 modes by using 25.9 M
$\psi^\prime$ events, and no evidence found. Product branching
fraction upper limits are determined as a function of
$\Gamma(\eta_c(2S))$ for the 11 individual
modes~\cite{cleo-c-etac-2s}.
\begin{figure}[htb]
  \begin{center}
    \includegraphics[width=0.4\textwidth]{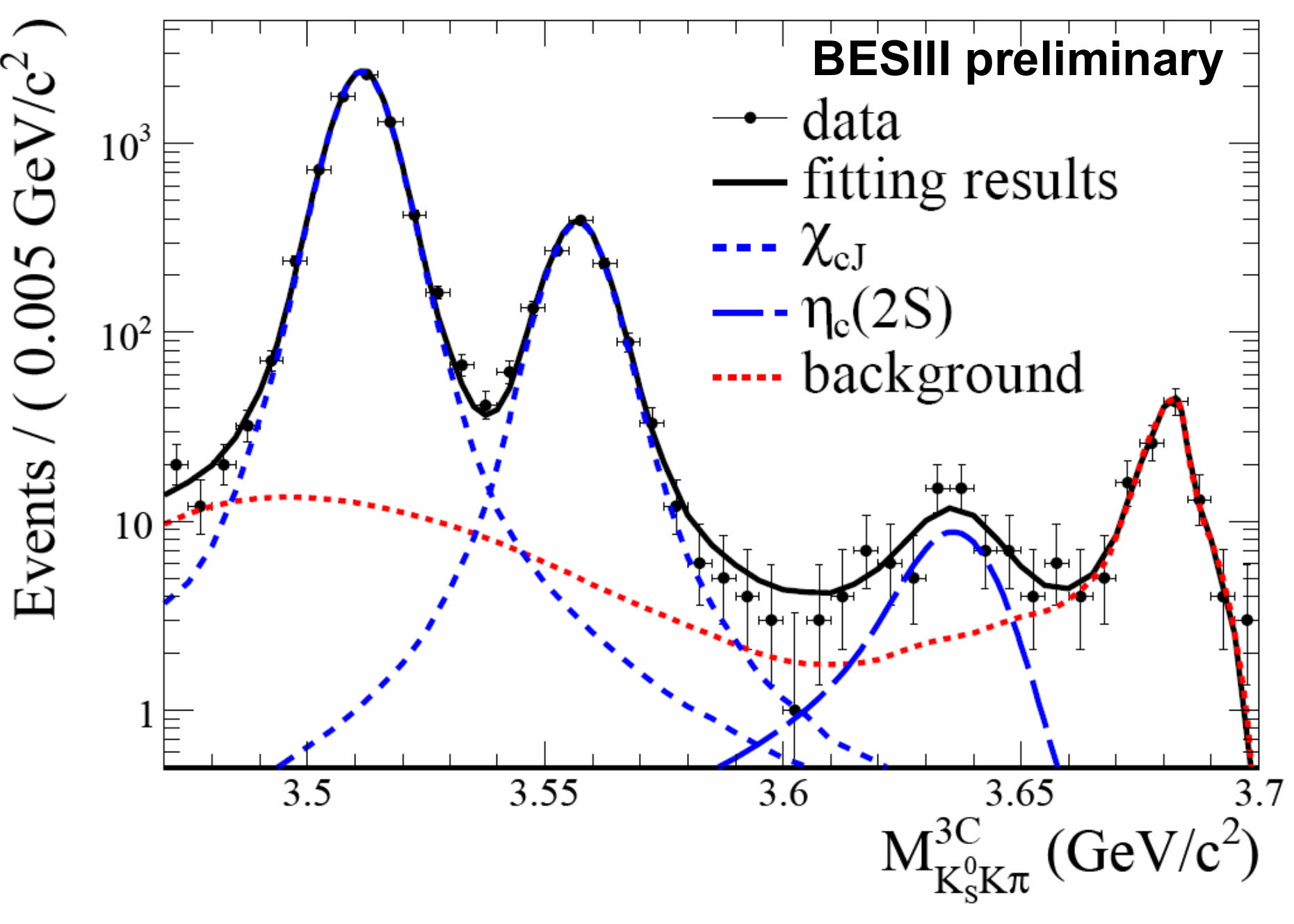}
    \caption{Fitting of the mass spectrum for $\eta_c(2S) \rightarrow K_s K^\pm \pi^\mp$.}
    \label{fig:etac-2s}
  \end{center}
\end{figure}

Using the largest $\psi^\prime$ data sample in the world which was
collected by the BESIII, we searched for the M1 transition
$\psi^\prime \rightarrow \gamma \eta_c(2S)$ through the hadronic
final states $K_sK^\pm\pi^\mp$. A bump is observed around 3635 MeV
on the mass spectrum as shown in Fig.~\ref{fig:etac-2s}. In order to
determine the background and mass resolution using data, the mass
spectrum range is enlarged (3.47 $\sim$ 3.72 GeV/c$^2$) to include
$\chi_{c1}$ and $\chi_{c2}$ events. The resonances $\chi_{c1}$ and
$\chi_{c2}$ are described by the corresponding Monte Carlo (MC)
shape convolved a Gaussian which takes account the small difference
on the mass shift and resolution between data and MC. So the mass
resolution for the $\eta_c(2S)$ in the fitting is fixed to the
linear extrapolation of the mass resolutions from the $\chi_{c1}$
and $\chi_{c2}$ signals in data. The line shape for $\eta_c(2S)$
produced by such the M1 transition is described by $(E_{\gamma}^3
\times BW(m)\times damping(E_\gamma)) \otimes Gauss(0,\sigma) $
where $m$ is the invariant mass of $K_s K^\pm \pi^\mp$, $E_\gamma =
\frac{m^2_{\psi^\prime}-m^2}{2m^2_{\psi^\prime}}$ is the energy of
the transition photon in the rest frame of $\psi^\prime$,
$damping(E_\gamma)$ is the function to damp the diverging tail
raised by $E^3_\gamma$ and $Gauss(0,\sigma)$ is the Gaussian
function describing the detector resolution. The possible form of
the damping function is somewhat arbitrary, and one suitable
function used by KEDR for a similar process is~\cite{kedr}
$$
\frac{E_0^2}{E_\gamma E_0 + (E_\gamma E_0 - E_0)^2}
$$
where $E_0
=\frac{m^2_{\psi^\prime}-m_{\eta_c(2S)}^2}{2m^2_{\psi^\prime}}$ is
the peaking energy of the transition photon. In the fit, the width
of $\eta_c(2S)$ is fixed to PDG value. From the fit to the data, a
signal with a statistical significance of 6.5 standard deviation is
observed which is the first observation of the M1 transition
$\psi^\prime \rightarrow \gamma \eta_c(2S)$. The measured mass for
$\eta_c(2S)$ is $3638.5 \pm 2.3 \pm 1.0$ MeV/$c^2$. The measured
branching ratio is $BR(\psi^\prime \rightarrow \gamma
\eta_c(2S))\times BR(\eta_c(2S) \rightarrow K_sK^\pm\pi^\mp) = (2.98
\pm 0.57\pm 0.48) \times 10^{-6}$. Together with the BABAR result
$BR(\eta_c(2S)\rightarrow K\bar{K}\pi) = (1.9 \pm 0.4
\pm1.1)\%$~\cite{babar-gg-1}, the M1 transition rate for
$\psi^\prime \rightarrow \gamma \eta_c(2S)$ is derived as
$BR(\psi^\prime \rightarrow \gamma \eta_c(2S))= (4.7 \pm 0.9 \pm
3.0)\times 10^{-4}$.

Most recently, the Belle Collaboration measured $\eta_c(1S)$ and
$\eta_c(2S)$ resonant parameters in the decay of $B^\pm \rightarrow
K^\pm (K_sK\pi)^0$ by considering the interference between
$\eta_c(1S)$/$\eta_c(2S)$ decay and non-resonant in the $B^\pm$
decay~\cite{belle-etac-update}. Meanwhile, the BABAR Collaboration
also updated the analysis of $e^+e^- \rightarrow
e^+e^-(\gamma\gamma), \gamma\gamma \rightarrow
\eta_c(1S)/\eta_c(2S)\rightarrow (K_sK\pi)^0$ and
$K^+K^-\pi^+\pi^-\pi^0$ modes~\cite{babar-etac-update}.
Table~\ref{table1} shows the summary of the typical production
processes ($\psi^\prime$ radiative decays, $\gamma\gamma$ fusion and
$B$ decays) for $\eta_c(1S)/\eta_c(2S)$ and corresponding resonant
parameter measurements. These results indicate that the $\eta_c(1S)$
parameters agree well from different production processes.

\begin{table}[tb]
{\small
  \begin{center}
    \begin{tabular}{lcccc}\hline
              & BESIII~\cite{this} & Belle~\cite{belle-etac-update} & BABAR~\cite{babar-etac-update} & PDG 2010~\cite{ref:hc-hyper}              \\
              & $\psi^\prime \rightarrow \gamma \eta_c/\eta_c(2S)$ &
              $B$ decays & $\gamma\gamma$ fusion &   \\
      \hline
      $M(\eta_c(1S))$ MeV/$c^2$   &$2984.4\pm0.5\pm0.6$ & $2985.4\pm1.5^{+0.2}_{-2.0}$ &  $2982.2\pm0.4\pm1.4$ & $2980.3\pm1.2$  \\
      $\Gamma(\eta_c(1S))$ MeV & $30.5\pm1.0\pm0.9$ & $35.1\pm3.1^{+1.0}_{-1.6}$ & $32.1\pm1.1\pm1.3$& $28.6\pm2.2$  \\
$M(\eta_c(2S))$ MeV/$c^2$ & $3638.5\pm2.3\pm1.0$ &
$3636.1^{+3.9+0.5}_{-1.5-2.0}$ & $3638.5\pm1.5\pm0.8$ & $3637\pm4$
\\
$\Gamma(\eta_c(2S))$ MeV & $12$ (fixed) &
$6.6^{+8.4+2.6}_{-5.1-0.9}$ & $13.4\pm4.6\pm3.2$ & $14\pm7$ \\
\hline
    \end{tabular}
    \caption{Comparison of the mass and width for $\eta_c(1S)/\eta_c(2S)$ in different production processes,
    $\psi^\prime\rightarrow \gamma \eta_c(1S)/\eta_c(2S)$, $B^\pm \rightarrow K^\pm \eta_c(1S)/\eta_c(2S)$ and
     $\gamma \gamma$ fusion, from different experiments. The PDG values are only world average from earlier results.
     . For the time being, the most precise measurements for $\eta_c(1S)$ resonance are from BESIII, while these for
     $\eta_c(2S)$ resonant parameters are from BABAR in $\gamma\gamma$ fusion.}
    \label{table1}
  \end{center}}
\end{table}

\subsection{$h_c(1P)$ properties}

The BESIII Collaboration reported the results on the production and
decay of the $h_c$ using 106M of $\psi^\prime$ decay events in
2010~\cite{hc_paper}, where we studied the distributions of mass
recoiling against a detected $\pi^0$ to measure $\psi^\prime
\rightarrow \pi^0 h_c$ both inclusively (E1-untagged) and in events
tagged as $h_c \rightarrow \gamma \eta_c$ (E1-tagged) by detection
of the E1 transition photon. In 2011, 16 specific decay modes of
$\eta_c$ are used to reconstruct $\eta_c$ candidates in the decay
mode of $h_c\rightarrow \gamma \eta_c$.
Figure~\ref{fig2:hc-exclusive} shows the $\pi^0$ recoiling mass for
individual $\eta_c$ decay modes in the decay chain of $\psi^\prime
\rightarrow \pi^0 h_c$, $h_c \rightarrow \gamma \eta_c$, while
Fig.~\ref{fig:hc-sum} is the sum of the 16 decay modes.  We fit the
16 $\pi^0$ recoil-mass spectra simultaneously that yields $M(h_c)=
3525.31\pm 0.11 (stat.)\pm0.15(syst.)$ MeV/$c^2$ and $\Gamma(h_c)=
0.70\pm0.28 (stat.)\pm0.25(syst.)$ MeV/$c^2$. These preliminary
results are consistent with the previous BESIII inclusive results
and CLEO-c exclusive results.

The centroid of the $^3P_J$ states
($\chi_{c0}$,$\chi_{c1}$,$\chi_{c2}$) is known to be $\langle
M(^3P_J)\rangle  = [5M(^3P_2) + 3M(^3P_1) + M(^3P_0)] = 3525.30 \pm
0.04$ MeV~\cite{ref:hc-hyper}. If the $^3P_J$ states centroid mass
$\langle M(^3P_J) \rangle$ is identified as the mass of $M(^3P)$,
then BESIII observes the hyperfine splitting as $\Delta
M_{hf}(1P)_{cc} = -0.01 \pm 0.11(stat.) \pm 0.14 (syst.)$ MeV which
agrees with zero.

\begin{figure}[htb]
  \begin{center}
    \includegraphics[width=0.7\textwidth]{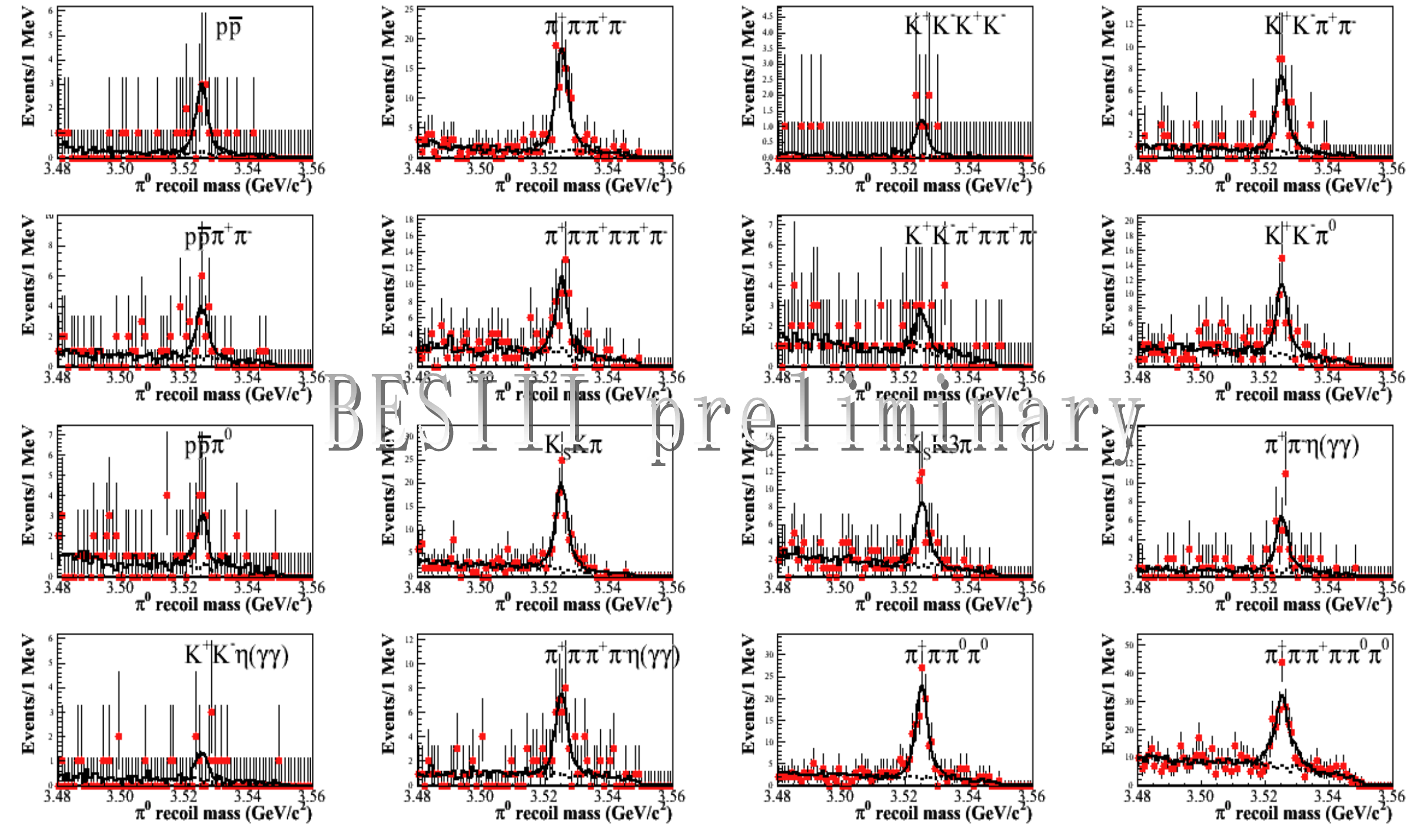}
    \caption{The $\pi^0$ recoiling mass for 16 $\eta_c$ decay modes.}
    \label{fig2:hc-exclusive}
  \end{center}
\end{figure}
\begin{figure}[htb]
  \begin{center}
    \includegraphics[width=0.4\textwidth]{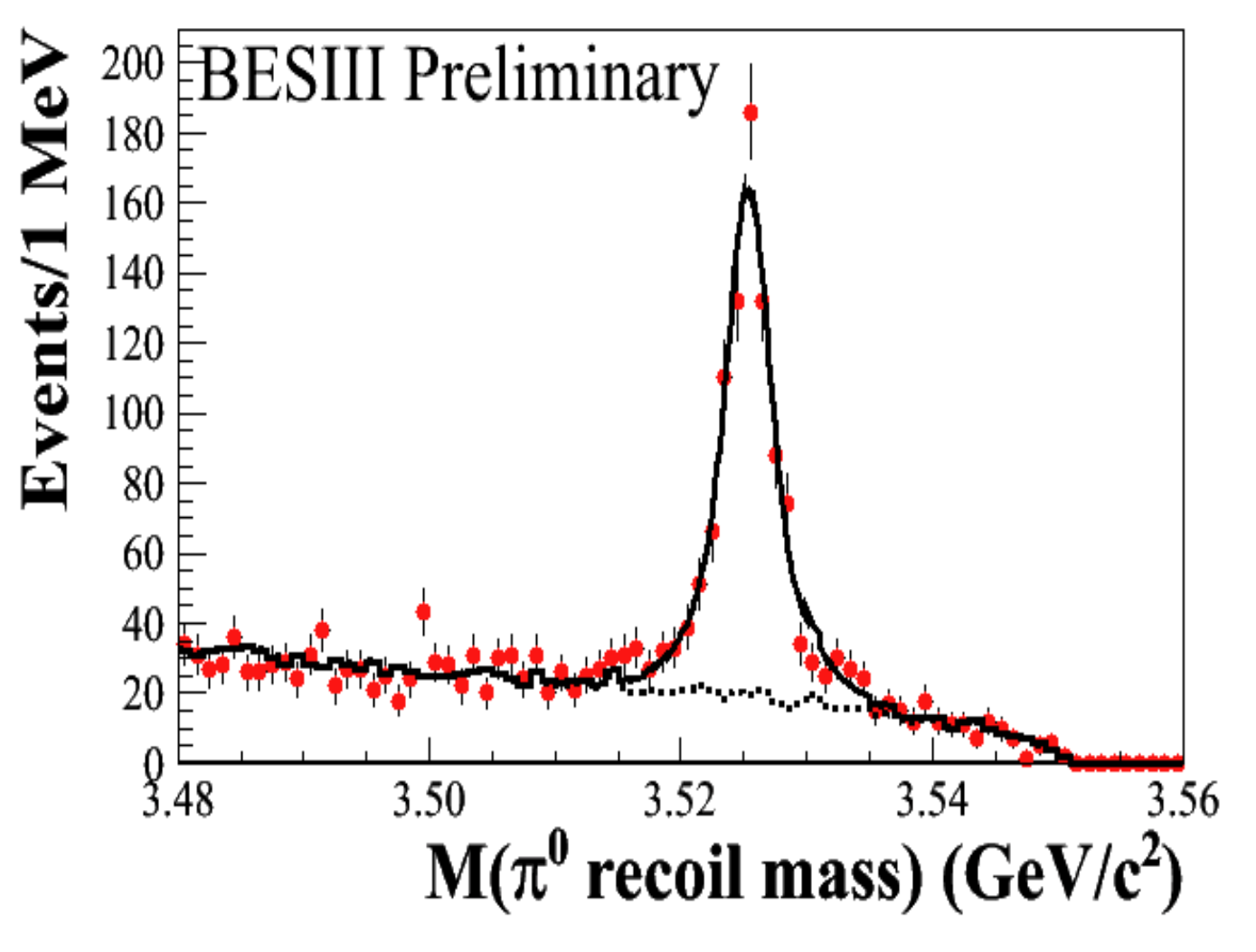}
    \caption{The $\pi^0$ recoiling mass for the sum of 16 $\eta_c$ decay modes.}
    \label{fig:hc-sum}
  \end{center}
\end{figure}

We also look at the $\eta_c(1S)$ mass distributions in the exclusive
decay modes. Figure~\ref{fig:etac-hc} shows the distribution of
invariant mass distribution from exclusive hadronic decays of
$\eta_c(1S)$, the shape of the $\eta_c$ is symmetric in $h_c(1P)
\rightarrow \gamma \eta_c(1S)$ radiative transition and no
distortion observed as in $J/\psi$/$\psi^\prime$ radiative decays.
The analysis is still on going for the resonant parameters.
\begin{figure}[htb]
  \begin{center}
    \includegraphics[width=0.6\textwidth]{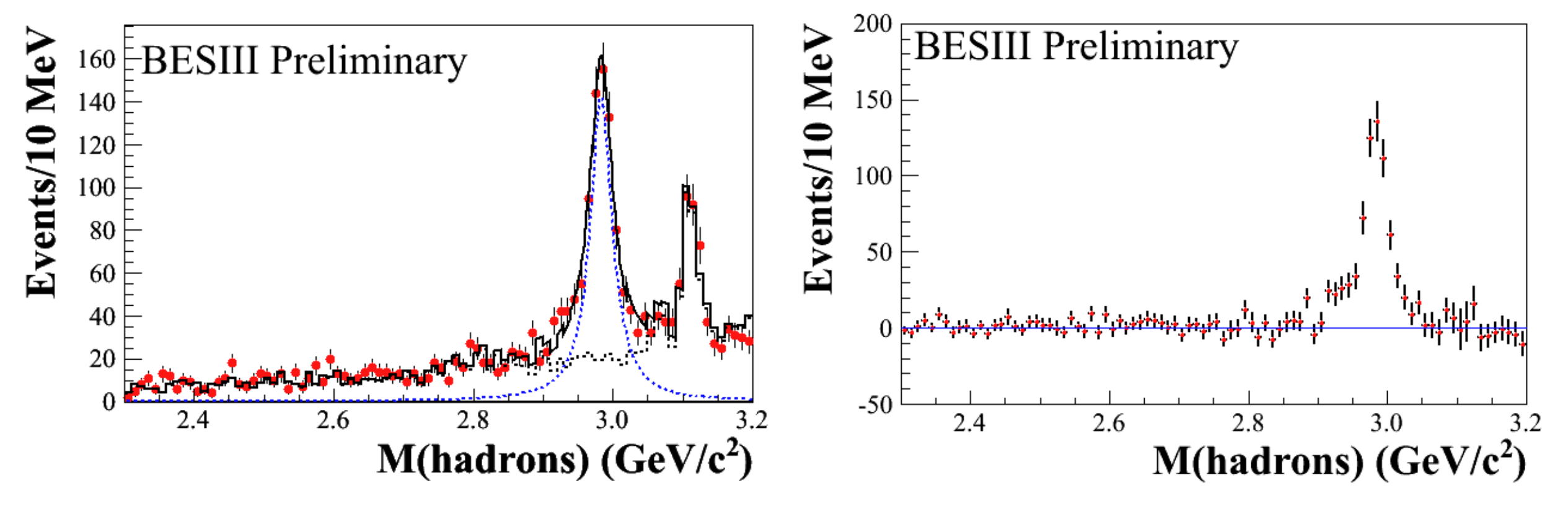}
    \caption{The invariant mass distributions for the sum of 16 $\eta_c$ decay modes. Right is background subtracted.}
    \label{fig:etac-hc}
  \end{center}
\end{figure}

\subsection{The radiative decays $\psi^\prime\rightarrow \gamma P$ with $P=\{\pi^0,\eta,\eta^\prime\}$}

The radiative decay of the $\psi^\prime$ to a pseudo-scalar meson,
such as the $\pi^0$, $\eta$, and $\eta^\prime$, is of interest since
it can be used to study the two-gluon coupling to $c\bar c$ states,
to study the $\eta - \eta^\prime$ mixing angle, and to probe the
$\pi^\circ$ form factor in the time-like region. Recently, the
CLEO-c Collaboration reported measurements for the decays of
$J/\psi$, $\psi^\prime$, and $\psi^{\prime\prime}$ to
 $\gamma P$~\cite{cleo_gP_paper}, and no evidence for $\psi^\prime\rightarrow \gamma\eta$ or
 $\gamma\pi^0$ was found.
With 106M $\psi^\prime$ data sample at BESIII, the processes
$\psi^\prime\rightarrow \gamma\pi^0$ and $\psi^\prime\rightarrow
\gamma\eta$ are observed for the first time with signal
significances of 4.6$\sigma$ and 4.3$\sigma$, respectively, and with
branching fractions of
$B(\psi^\prime\rightarrow\gamma\pi^0)=(1.58\pm0.40\pm0.13)\times
10^{-6}$ and $B(\psi^\prime\rightarrow
\gamma\eta)=(1.38\pm0.48\pm0.09)\times 10^{-6}$. With a measured
branching fraction,
$B(\psi^\prime\rightarrow\gamma\eta^\prime)=(126\pm 3\pm 8)\times
10^{-6}$. The mass distributions of the pseudo-scalar meson
candidates can be found in Fig.~\ref{fig:psip-gp} the BESIII
Collaboration determined for the first time the ratio of the $\eta$
and $\eta^\prime$ production rates from $\psi^\prime$ decays,
$R_{\psi^\prime}\equiv B(\psi^\prime\rightarrow
\gamma\eta)/B(\psi^\prime\rightarrow
\gamma\eta^\prime)=(1.10\pm0.38\pm0.07)$\%. This ratio is below the
90\% C.L. upper bound determined by the CLEO-c Collaboration and, in
contradiction to predictions of leading-order perturbative-QCD, one
order of magnitude smaller than the corresponding ratio for the
$J/\psi$ decays, $R_{J/\psi}=(21.1\pm0.9)$\%. More details on the
data analysis can be found in Ref.~\cite{gP_paper}. These
observations are interpreted in the framework of vector meson
dominance (VMD) in association with the $\eta_c - \eta(
\eta^\prime$) mixings due to the axial gluonic
anomaly~\cite{zhao-gammap}.

\begin{figure}[htb]
  \begin{center}
    \includegraphics[width=0.6\textwidth]{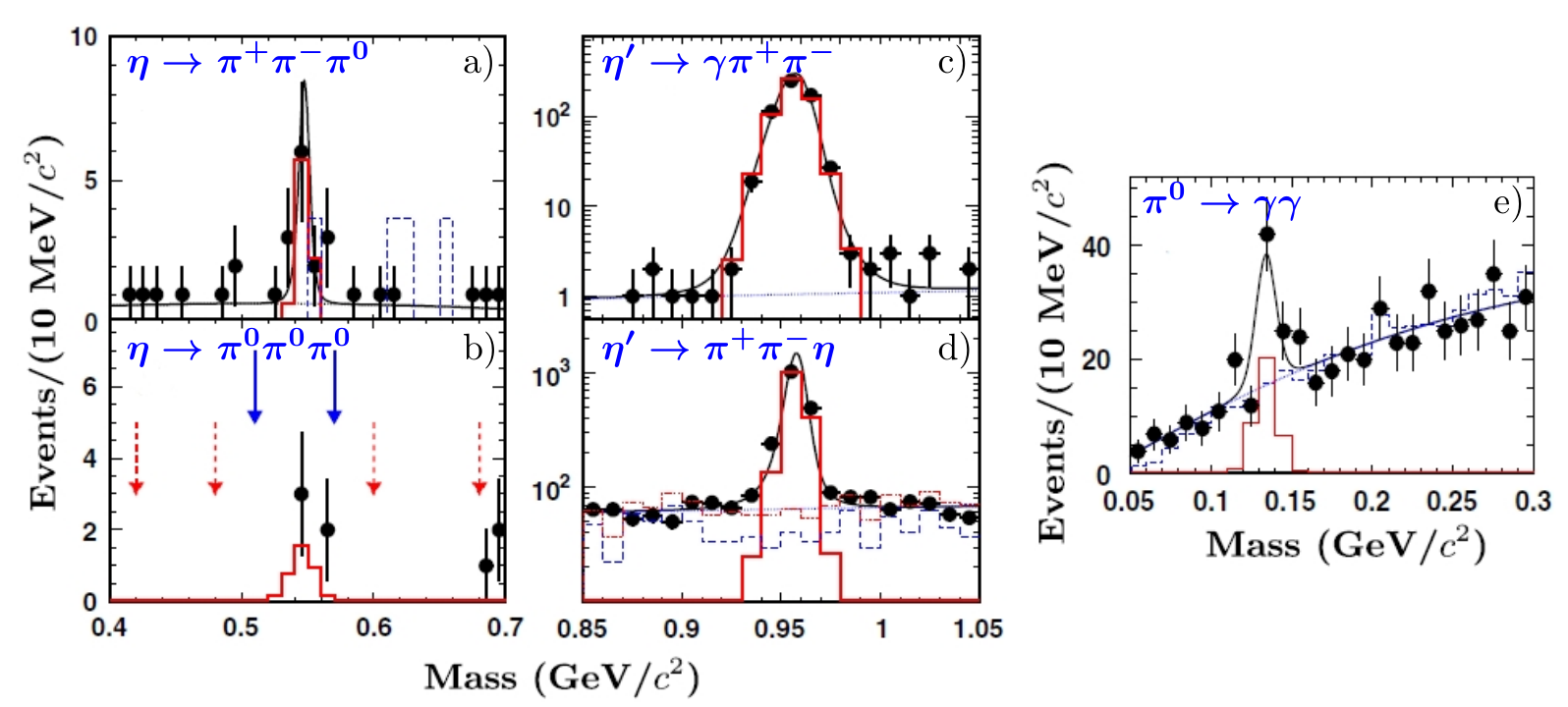}
    \caption{Mass distributions of the pseudo-scalar meson candidates for $\psi^\prime\rightarrow \gamma P$:
a) $P=\eta(\rightarrow\pi^+\pi^-\pi^0)$; b) $P=\eta(\rightarrow
3\pi^0)$; c) $P=\eta^\prime(\rightarrow \gamma\pi^+\pi^-)$; d)
$P=\eta^\prime(\rightarrow \pi^+\pi^-\eta(\rightarrow
\gamma\gamma))$; d) $P=\pi^\circ(\rightarrow \gamma\gamma)$. For
more details, we refer to Ref.~\cite{gP_paper}.}
    \label{fig:psip-gp}
  \end{center}
\end{figure}

\subsection{Measurements of $J/\psi \rightarrow N\bar{N}$ branching fractions}

\begin{figure}[htb]
  \begin{center}
    \includegraphics[width=0.3\textwidth]{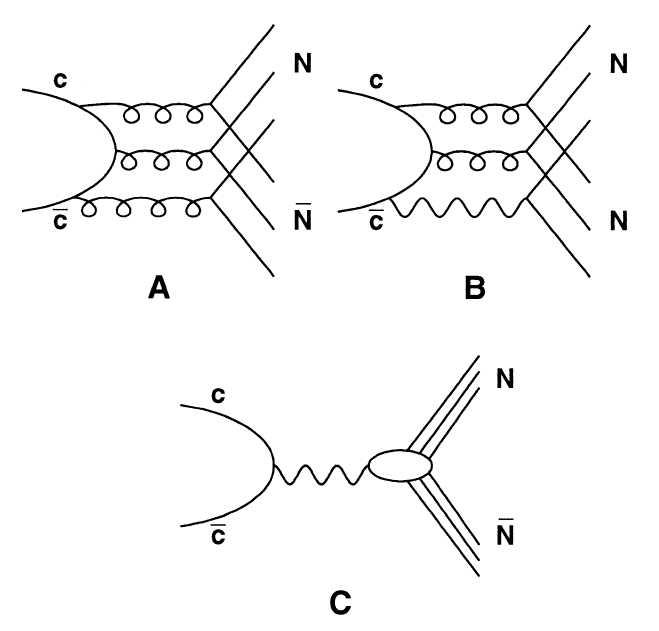}
    \caption{Feynman diagrams of the three main contributions to the
J/$\psi$ decay in nucleon-antinucleon. A) purely strong amplitude,
B) strong-electromagnetic amplitude and C) purely electromagnetic
amplitude. These plots are from reference~\cite{ppbar-fen}}
    \label{fig:ppbar}
  \end{center}
\end{figure}

The expected three main contributions to the decay $J/\psi
\rightarrow N\bar{N}$ are shown in Fig.~\ref{fig:ppbar}. This decay
mode should be a very good test of PQCD, because of the 3 gluons in
the OZI violating J/$\psi$ strong decay just matching the 3
($q\bar{q}$) pairs (as in Fig.~\ref{fig:ppbar}(a)). The ratio
between $BR(J/\psi \rightarrow n\bar{n})$ and $BR(J/\psi \rightarrow
p\bar{p})$ strongly depends on the phase between the strong and the
electromagnetic (e.m.) contributions as shown in
Fig.~\ref{fig:ppbar}~\cite{ppbar-fen}. According to PDG values,
$BR(J/\psi \rightarrow n\bar{n})$ is very similar to $BR(J/\psi
\rightarrow p\bar{p})$, which gives a clear indication that strong
and e.m. contributions do not interfere and have to be added in
quadrature. An estimate for the relative phase $\phi$ on the basis
of previous experimental results is $\phi =
89^o\pm15^o$~\cite{ppbar-fen}. However, the current experimental
uncertainty on $BR(J/\psi \rightarrow n\bar{n})$ is still large
[($2.2\pm0.4)\times 10^{-3}$], which is main contribution to the
uncertainty on the phase determination. Therefore, it motivates the
BESIII experiment to measure $J/\psi \rightarrow n\bar{n}$ decay
rate with high statistical $J/\psi$ sample.

Based on 225M J/$\psi$ events collected at BESIII, the anti-neutron
is identified by shower energy deposition and hit information in
EMC, with constraints of known center-of-mass energy and
back-to-back between neutron and anti-neutron, the branching ratio
is determined to be $BR(J/\psi \rightarrow n\bar{n})= (2.07 \pm 0.01
\pm 0.17)\times 10^{-3}$. At the same time, the branching ratio for
$ J/\psi \rightarrow p\bar{p}$ is determined to be $BR(J/\psi
\rightarrow p\bar{p}) = (2.112 \pm 0.004 \pm 0.027)\times 10^{-3}$.
These result improve by a large factor comparing to the previous
measurements and strongly support the orthogonal phase of strong and
e.m. amplitudes.

\subsection{$\eta^\prime$ and $\eta(1405)$ in J/$\psi \rightarrow
\gamma \pi\pi\pi$ decays}

\begin{figure}[htb]
  \begin{center}
    \includegraphics[width=1.0\textwidth]{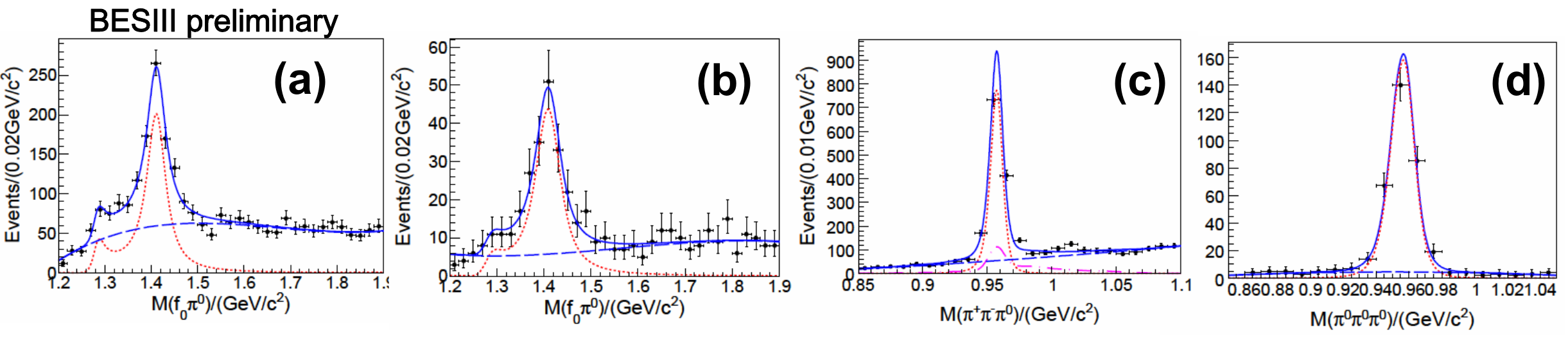}
    \caption{The invariant mass distributions:
    (a) invariant mass of $f_0(980) \pi^0$ from J/$\psi \rightarrow \gamma \pi^+\pi^-\pi^0$; (b) invariant mass of $f_0(980) \pi^0$ mass
    from J/$\psi \rightarrow \gamma \pi^0\pi^0\pi^0$;  (c) the invariant mass of $\eta^\prime \rightarrow 3\pi$s in
     J/$\psi \rightarrow \gamma \pi^+\pi^-\pi^0$, (d) mass of $\eta^\prime \rightarrow 3\pi$s from J/$\psi \rightarrow \gamma \pi^0\pi^0\pi^0$, respectively.}
    \label{fig:eta1405}
  \end{center}
\end{figure}

The spectrum of radial excitation states of isoscalar $\eta$ and
$\eta^\prime$ is still not well known. An important issue is about
the nature of $\eta(1405)$ and $\eta(1475)$ states, which are not
well established. BESIII measured the decays of J/$\psi \rightarrow
\gamma \pi^+\pi^-\pi^0$ and $\gamma \pi^0\pi^0\pi^0$. In the two
decay modes, clear $f_0(980)$ signals are observed on both
$\pi^+\pi^-$ and $\pi^0\pi^0$ spectra, the width of observed
$f_0(980)$ is much narrower ($\sim$ 10 MeV) than that in other
processes~\cite{ref:hc-hyper}. By taking events in the window of
$f_0(980)$ on the $\pi\pi$ mass spectrum, we observed evidence of
$f_1(1285)$ in the low mass region of $f_0(980) \pi^0$ as shown in
Fig.~\ref{fig:eta1405} (a) and (b), which corresponding to
significance of about 4.8 $\sigma$ for $f_1(1285) \rightarrow
f_0(980)\pi^0$ in $f_0(980) \rightarrow \pi^+\pi^-$ mode (1.4
$\sigma$ in $f_0(980) \rightarrow \pi^0\pi^0$ ). It is interesting
that clear peak around 1400 MeV is also observed on the mass of
$f_0(980) \pi^0$ (see Fig.~\ref{fig:eta1405} (a) and (b)).
Preliminary angular analysis indicates that the peak on 1400 MeV is
from $\eta(1405) \rightarrow f_0(980) \pi^0$ decay. BESIII measured
the combined branching fraction of $\eta(1405)$ production to be
$BR(J/\psi \rightarrow \gamma \eta(1405) )\times BR(\eta(1405)
\rightarrow f_0(980) \pi^0)\times BR(f_0(980) \rightarrow \pi^+
\pi^-) = (1.48 \pm 0.13 \pm 0.17)\times 10^{-5}$ and $BR(J/\psi
\rightarrow \gamma \eta(1405))\times BR(\eta(1405) \rightarrow
f_0(980) \pi^0) \times BR(f_0(980) \rightarrow \pi^0\pi^0) = (6.99
\pm 0.93 \pm 0.95)\times 10^{-6}$, respectively. It is the first
time that we observe anomalously large isospin violation in the
strong decay of $\eta(1405) \rightarrow f_0(980) \pi^0$.

Following BESIII measurements, in reference~\cite{zhaoq}, the
authors interpret this puzzle as an intermediate on-shell
$K\bar{K^*} + c.c.$ rescattering to the isospin violating
$f_0(980)\pi^0$ by exchanging on-shell kaon. Further experimental
study on the $\eta(1405)$ parameters and identification of quantum
number are needed at BESIII.

By looking at the invariant mass of $\pi\pi\pi$ in $J/\psi
\rightarrow \gamma 3\pi$s decays as shown in Fig.~\ref{fig:eta1405}
(c) and (d), we observe signals for $\eta^\prime \rightarrow
\pi^+\pi^- \pi^0$ and $\pi^0\pi^0\pi^0$ decays, respectively,  and
determine the decay rates to be $BR(\eta^\prime \rightarrow
\pi^+\pi^- \pi^0) = (3.83\pm 0.15\pm 0.39)\times 10^{-3}$ and
$BR(\eta^\prime \rightarrow \pi^0\pi^0 \pi^0) = (3.56\pm
0.22\pm0.34)\times 10^{-3}$, respectively. For $\eta^\prime
\rightarrow \pi^+\pi^- \pi^0$ decay, it is consistent with CLEO-c's
measurements and precision is improved by a factor of 4. In contrast
, for $\eta^\prime \rightarrow \pi^0\pi^0 \pi^0$ decay, it is two
times larger than that in the PDG value~\cite{ref:hc-hyper}.

\subsection{Observation of resonances above 2.0 GeV
in $J/\psi \rightarrow \gamma \eta^\prime \pi^+\pi^-$ decay}

\begin{figure}[htb]
  \begin{center}
    \includegraphics[width=1.0\textwidth]{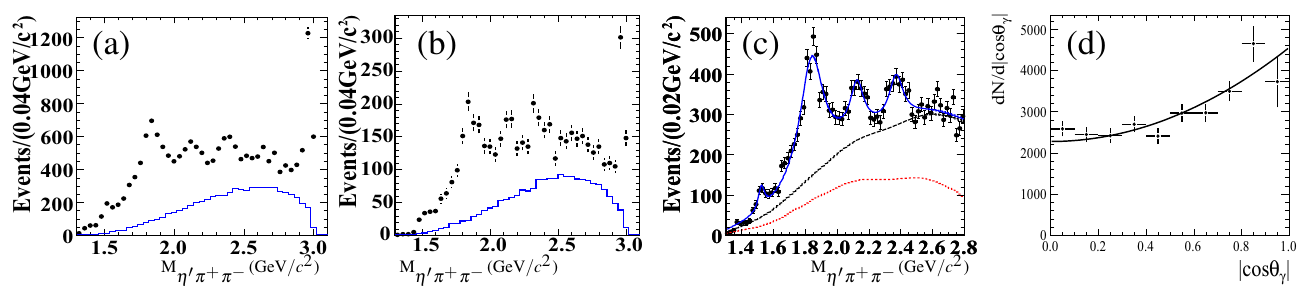}
    \caption{The invariant mass distributions of $\eta^\prime
    \pi^+\pi^-$ with (a) $\eta^\prime \rightarrow \gamma \rho$ (b)
    $\eta^\prime \rightarrow \eta \pi^+\pi^-$, and (c) the
    combined plot for (a) and (b), and fitting results. (d) is for
    the cos($\theta_\gamma$) distribution, where $\theta_\gamma$ is
    the pole angle of the radiative photon. The points with error
    bar are data. The histograms are from $J/\psi \rightarrow \gamma \eta^\prime \pi^+\pi^-$ phase space MC.}
    \label{fig:x1835}
  \end{center}
\end{figure}

The $X(1835)$ was firstly observed at the BESII with a statistical
significance of 7.7$\sigma$~\cite{bes2:x1835}. The possible
interpretations of the $X(1835)$ include a $p\bar{p}$ bound
state~\cite{x1835-ppbar1,x1835-ppbar2}, a
glueball~\cite{x1835-glueball}, a radial excitation of the
$\eta^\prime$ meson~\cite{x1835-eta}, etc. We report a study of
J/$\psi \rightarrow \gamma \eta^\prime \pi^+\pi^-$  that used two
$\eta^\prime$ decay modes, $\eta^\prime \rightarrow \gamma \rho$ and
$\eta^\prime \rightarrow \pi^+\pi^- \eta$.  Figure~\ref{fig:x1835}
(a) and (b) show the $\eta^\prime \pi^+\pi^-$ invariant mass
spectrum in J/$\psi \rightarrow \gamma \eta^\prime \pi^+\pi^-$ with
$\eta^\prime \rightarrow \gamma \rho$ and $\eta^\prime \rightarrow
\pi^+\pi^- \eta$ decay modes, respectively. The $X(1835)$ resonance
is clearly seen. Additional peaks are observed around 2.1 and 2.3
GeV/c$^2$, denoted as $X(2120)$ and $X(2370)$, as well as the
$f_1(1510)$ in the low mass region and the distinct $\eta_c(1S)$
signal in the high mass region.

Fits to the mass spectra have been made using four
 efficiency-corrected
Breit-Wigner functions convolved with a Gaussian mass resolution
plus a nonresonant $\pi^+\pi^- \eta^\prime$ contribution and
background representations. The fitting result of the combined mass
spectrum is shown in Fig.~\ref{fig:x1835}(c).  The mass and width of
X(1835) are measured to be $M=1836.5 \pm 3.0^{+5.6}_{-2.1}$
MeV/c$^2$ and $\Gamma = 190 \pm 9^{+38}{-36}$ MeV with a
significance larger than 20$\sigma$. The masse and width for
$X(2120)$ ($X(2370)$) is determined to be $M= 2122.4 \pm
6.7^{+4.7}_{-2.7}$ MeV/c$^2$ ($M=2376.3\pm 8.7^{+3.2}_{-4.3}$
MeV/c$^2$) and $\Gamma = 83\pm 16^{+31}_{-11}$ MeV ($\Gamma = 83\pm
17^{+44}_{-6}$ MeV) with significance 7.2$\sigma$(6.4$\sigma$). For
$X(1835)$, the $cos\theta_\gamma$ distribution is shown in
Fig.~\ref{fig:x1835}, where $\theta_\gamma$ is the polar angle of
the radiative photon in the J/$\psi$ center of mass system. It
agrees with $(1+cos^2\theta_\gamma)$, which is expected for a
pseudoscalar.

\subsection{Observation of $X(1870)\rightarrow a_0^\pm \pi^\mp $ in $J/\psi \rightarrow \omega
\eta\pi^+\pi^-$ decay}
\begin{figure}[htb]
  \begin{center}
    \includegraphics[width=0.7\textwidth]{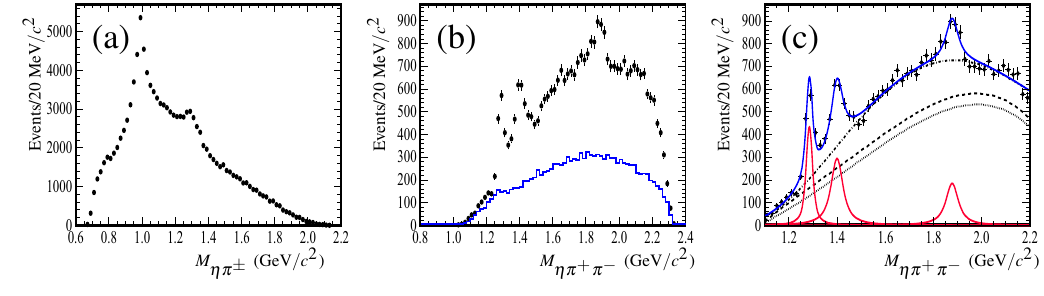}
    \caption{The invariant mass distributions of (a) $\eta\pi^\pm$, (b) $\eta \pi^+\pi^-$ with events in $a_0(980)$ mass
    window;
    (c): fitting results;  In (b), the histograms are from $J/\psi \rightarrow \omega \eta \pi^+\pi^-$ phase space MC.}
    \label{fig:x1870}
  \end{center}
\end{figure}

Experimentally, the study of the production mechanism of the
$X(1835)$ and $\eta1405)$, e.g. searches for them in $\eta \pi\pi$
final states with other accompanying particles ($\phi$, $\omega$
etc.), are useful for clarifying their nature. In particular, the
measurements of the production width of these two states in hadronic
decays of the J/$\psi$ and a comparison with corresponding
measurements in J/$\psi$ radiative decay would provide important
information about the glueball possibility. We present results of a
study of J/$\psi \rightarrow \omega \eta \pi^+\pi^-$, in which the
$\omega$ decays to $\pi^+\pi^-\pi^0$ and the $\eta /\pi^0$ decays to
a pair of photons. In the mass spectrum of $\eta \pi^pm$, as shown
in Fig.~\ref{fig:x1870} (a), the $a^\pm_0(980)$ peak is clearly
seen. The mass spectrum of $\eta \pi^+\pi^-$ with events in the
$a_0(980)$ mass window is shown in Fig.~\ref{fig:x1870} (b). Both
$f_1(1285)$ and $\eta(1405)$ are observed significantly. A clear
peak around 1800 MeV, denoted as $X(1870)$ is also seen for the
first time in the decay of J/$\psi \rightarrow \omega \eta
\pi^+\pi^-$. A fit with three resonances with simple BW formula
yields a mass $M=1877.3 \pm 6.3^{+3.4}_{-7.4}$ MeV/c$^2$ and a width
$\Gamma = 57\pm 12^{+19}_{-4}$ MeV for the $X(1870)$ structure with
s statistical significance of 7.2$\sigma$. Whether the resonant
structure of $X(1870)$ is due to the $X(1835)$, the $\eta_2(1870)$,
 or a new resonance still needs further study such as a partial
wave analysis that will be possible with the larger J/$\psi$ data
sample. For the $\eta(1405)$, the product branching ratio of its
hadronic production is measured to be smaller than that for its
production in the radiative J/$\psi$ decays~\cite{ref:hc-hyper}. For
detail about this analysis, one can find more in
reference~\cite{bes3:x1870}.

\subsection{Open charm physics}
\begin{figure}[htb]
  \begin{center}
    \includegraphics[width=0.5\textwidth]{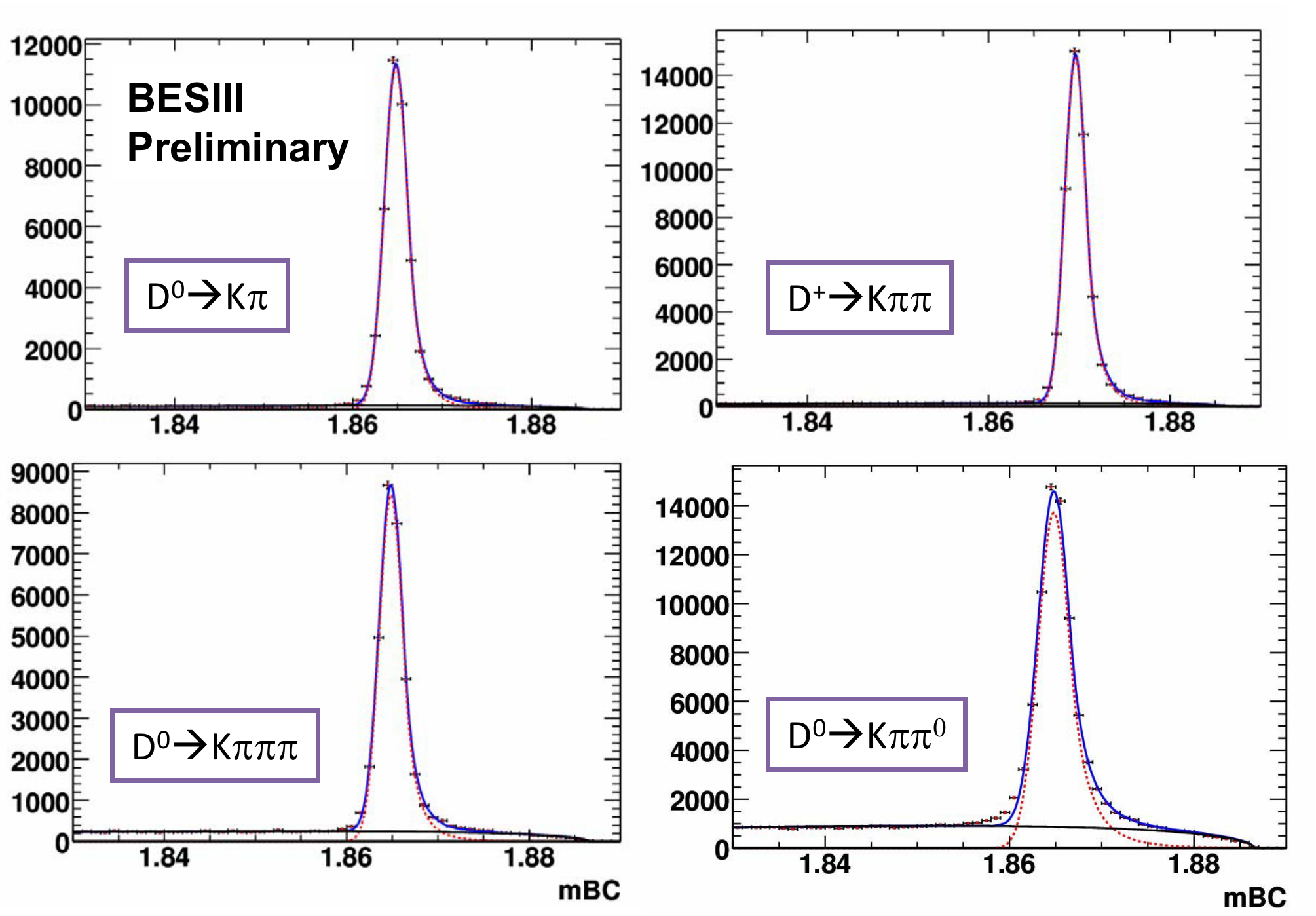}
    \caption{Beam constraint mass distributions.}
    \label{fig:dtag}
  \end{center}
\end{figure}

As of May, 2011, the BESIII has accumulated integrated luminosity of
2.9 fb$^{-f}$ on $\psi(3770)$ peak for open charm physics, which is
about 3.5 times the previous largest $\psi(3770)$ data set taken by
the CLEO-c Experiment. Taking advantage of the high luminosity
provided by the BEPCII collider, BESIII is expected to take much
more data at both 3770 MeV and higher energies. Many high precision
measurements, including CKM matrix elements related to charm weak
decays, decay constants $f_D$ and $f_{Ds}$, form factors from D
semileptonic decays, Dalitz decays, searches for $CP$ violation, and
absolute decay branching fractions, will be accomplished in the near
future.

There are many advantages of doing charm physics at experiments with
$e^+e^-$ collider at threshold. With $e^+e^-$ colliding, the initial
energy and quantum numbers are known. For $\psi(3770) \rightarrow
D\bar{D}$, both D daughter mesons can be fully reconstructed,
allowing absolute measurements, namely, the so called double-tag
technique. The continuum background is greatly suppressed by doing
this, and kinematic constraints can be applied to infer missing
particles on the other side D meson. To demonstrate the cleanliness
of the D tags, in Fig.~\ref{fig:dtag}, we show a beam-constrained
mass plots for some od the fully reconstructed $D^0/D^\pm$ decay
modes, from part of data. The beam-constrained mass $m_{BC} =
\sqrt{E^2_{beam}- p^2_D}$ , where $E_{beam}$ is the beam energy and
$p_D$ is the candidate momentum,  is obtained by substituting the
reconstructed $D$ energy with the beam energy.  With this technique,
the neutrino can be reconstructed in the leptonic and semileptonic
$D$ decays, so that decay constants and form factor/CKM matrix
elements can be precisely measured. Thus, these precise measurements
could be used to test and calibrate the theoretical tools such as
lattice QCD, which are critical for dealing with B decays.

\section{Summary}
The BESIII experiment addresses a wide range of topics in the field
of QCD and searches for new physics beyond the standard model.  At
present the BESIII collaboration has collected a record on
statistics on J/$\psi$, $\psi^\prime$, and $\psi(3770)$ charmonium
states. These data are being exploited to provide precision
measurements with a high discovery potential in light hadron and
charmonium spectroscopy, charmonium decays, and open charm
productions. Most recent highlights from BESIII are reviewed in this
paper.

Since very recently, data about 470 pb$^{-1}$ luminosity  have been
taken at a center-of-mass energy of 4010 GeV, which could give new
insights in our understanding of the recently discovered $XYZ$
states and which will allow to explore the field of $D_s$ physics.

\acknowledgements{%
We would like to thank the accelerator people at BEPCII for their
hard work which makes our high luminosity possible. We would also
like to thank the great computational and software support of the
IHEP staff. This work is supported in part by the Ministry of
Science and Technology of China under Contract No. 2009CB825200; 100
Talents Program of CAS.}


%

}  



\begin{thebibliography}{99}

\bibitem{besiii} M.~Ablikim {\it et al.} (BESIII Collaboration), \emph{Design andconstructionoftheBESIIIdetector},Nucl. Instrum. Meth. {\bf A614}, 345(2009).

\bibitem{hc_paper} M.~Ablikim {\it et al.} (BESIII Collaboration), \emph{Measurements of $h_c(^1P_1)$ in $\psi^\prime$ Decays}, Phys. Rev. Lett. {\bf 104}, 132002 (2010) [{\tt hep-ex/1002.0501}].
\bibitem{gP_paper} M.~Ablikim {\it et al.} (BESIII Collaboration), \emph{Evidence for $\psi^\prime$ decays into $\gamma\pi^0$ and $\gamma\eta$}, Phys. Rev. Lett. {\bf 105}, 261801 (2010) [{\tt hep-ex/1011.0885}].
\bibitem{PP_paper} M.~Ablikim {\it et al.} (BESIII Collaboration), \emph{Branching fraction measurements of $\chi_{c0}$ and $\chi_{c2}$ to $\pi^0\pi^0$ and $\eta\eta$},  Phys. Rev. D {\bf 81}, 052005 (2010) [{\tt hep-ex/1001.5360}].
\bibitem{ppbar_paper} M.~Ablikim {\it et al.} (BESIII Collaboration), \emph{Observation of a $p\bar p$ mass threshold enhancement in $\psi^\prime\rightarrow \pi^+ \pi^- J/\psi (J/\psi\rightarrow \gamma p\bar p)$ decay}, Chin. Phys. C {\bf 34}, 4 (2010) [{\tt hep-ex/1001.5328}].
\bibitem{X1835_paper} M.~Ablikim {\it et al.} (BESIII Collaboration), \emph{Confirmation of the $X(1835)$ and observation of the resonances $X(2120)$ and $X(2370)$ in $J/\psi\to \gamma \pi^+\pi^-\eta^\prime$}, Phys. Rev. Lett. {\bf 106}, 072002 (2011) [{\tt hep-ex/1012.3510}].
\bibitem{matrix_paper} M.~Ablikim {\it et al.} (BESIII Collaboration), \emph{Measurement of the Matrix Element for the Decay $\eta^{\prime} \to \eta \pi^+\pi^-$}, Phys. Rev. D {\bf 83}, 012003 (2011) [{\tt hep-ex/1012.1117}].
\bibitem{4pi_paper} M.~Ablikim {\it et al.} (BESIII Collaboration), \emph{First Observation of the Decays $\chi_{cJ} \to \pi^0 \pi^0 \pi^0 \pi^0$}, Phys. Rev. D {\bf 83}, 012006 (2011) [{\tt hep-ex/1011.6556}].
\bibitem{mixing_paper} M.~Ablikim {\it et al.} (BESIII Collaboration), \emph{Study of $a_0^0(980) - f_0(980)$ mixing}, Phys. Rev. D {\bf 83}, 032003 (2011) [{\tt hep-ex/1012.5131}].
\bibitem{gV_paper} M.~Ablikim {\it et al.} (BESIII Collaboration), \emph{Study of $\chi_{cJ}$ radiative decays into a vector meson}, Phys. Rev. D {\bf 83}, 112005 (2011) [{\tt hep-ex/1103.5564}].
\bibitem{vv_paper} M.~Ablikim {\it et al.} (BESIII Collaboration), \emph{Observation of $\chi_{c1}$ Decays into Vector Meson Pairs $\phi\phi$, $\omega\omega$, and $\omega\phi$}, Phys. Rev. Lett. {\bf 107}, 092001 (2011).


\bibitem{ref:hc-hyper}
 PDG 2010: Particle Data Group, \emph{Review of Particle Physics},
 Journal of Physics {\bf G 37}, 075021 (2010).

\bibitem{mark3} R. M. Baltrusaitis {\it et al.} (Mark-III Collaboration), Phys. Rev. {\bf D
33}, 629 (1986).

\bibitem{bes1} J. Z. Bai {\it et al.} (BES Collaboration), Phys. Lett. {\bf B 555}, 174
(2003).

\bibitem{cleo-c}D. M. Asner {\it et al.} (CLEO Collaboration), Phys. Rev. Lett. {\bf 92},
142001 (2004).
\bibitem{babar} B. Aubert {\it et al.} (BABAR Collaboration), Phys.
Rev. Lett. {\bf 92}, 142002 (2004).

\bibitem{belle1} S. Uehara {\it et al.} (Belle Collaboration), Eur. Phys. J. {\bf C 53}, 1
(2008).

\bibitem{belle2} A. Vinokurova {\it et al.} (Belle Collaboration), arXiv:1105.0978
[hep-ex].

\bibitem{cleo-c-new} R. E. Mitchell {\it et al.} (CLEO Collaboration), Phys.
Rev. Lett. {\bf 102}, 011801 (2009).
\bibitem{kedr} V.~V. Anashin {\it et al.} (KEDR
Collaboration), arXiv:1012.1694.
\bibitem{kseth} E. Eichten,  K. Gottfried, T. Kinoshita, K. D. Lane and T.-M. Yan, Phys.
Rev. {\bf D17}, 3090 (1978).

\bibitem{belle-etac-2s} S.-K. Choi {\it et al.} (Belle Collaboration), Phys. Rev. Lett. {\bf 89},
102001 (2002).
\bibitem{cleo-etac-2s} D. M. Asner {\it et al.} (CLEO Collaboration), Phys.
Rev. Lett. {\bf 92}, 142001 (2004).
\bibitem{babar-etac-2s} B. Aubert {\it et al.} (BABAR
Collaboration), Phys. Rev. Lett. {\bf 92}, 142002 (2004).
\bibitem{babar-etac-2s-double}  B. Aubert {\it et al.} (BABAR Collaboration), Phys. Rev. {\bf D 72}, 031101 (2005).

\bibitem{crystal-ball-etac-2s} C. Edwards {\it et al.} (Crystal Ball Collaboration), Phys. Rev. Lett.
{\bf 48}, 70 (1982).
\bibitem{cleo-c-etac-2s} D. Cronin-Hennessy {\it et al.} (CLEO Collaboration), Phys. Rev. {\bf D 81}, 052002
(2010).

\bibitem{babar-gg-1} B. Aubert {\it et al.} (Babar Collaboration), Phys. Rev. {\bf D 78},
012006 (2008).
\bibitem{this} L.~L.~Wang, for the BESIII Collaboration, this conference (preliminary results).
\bibitem{belle-etac-update} A. Vinokurova {\it et al.} (Belle Collaboration), arXiv:1105.0978.
\bibitem{babar-etac-update} P. del Amo Sanchez {\it et al.} (BABAR Collaboration), Phys. Rev. {\bf D 84}, 012004 (2011).
\bibitem{cleo_gP_paper} T. K. Pedlar {\it et al.} (CLEO Collaboration), Phys. Rev. D {\bf 79}, 111101 (2009).

\bibitem{zhao-gammap}
  Q.~Zhao,
  Phys.\ Lett.\  {\bf B697}, 52(2011).

\bibitem{ppbar-fen} R. Baldini {\it et al.} (FENICE Collaboration), Phys. Lett. {\bf B 444}, 111 (1998).
\bibitem{zhaoq} J.-J. Wu, X.-H. Liu, Q. Zhao and B.-S. Zou, arXiv:
1108.3772.
\bibitem{bes2:x1835} M.~Ablikim {\it et al.} (BESII Collaboration)
Phys. Rev. Lett. {\bf 95}, 262001 (2005).
\bibitem{x1835-ppbar1} G. J. Ding and M. L. Yan, Phys. Rev. {\bf C 72}, 015208 (2005).
\bibitem{x1835-ppbar2} G. J. Ding and M. L. Yan, Eur. Phys. J. {\bf A 28}, 351 (2006).
\bibitem{x1835-glueball} B. A. Li, Phys. Rev. D 74, 034019 (2006).
\bibitem{x1835-eta} T. Huang and S. L. Zhu, Phys. Rev. D 73, 014023 (2006).
 \bibitem{bes3:x1870} M.~Ablikim {\it et al.} (BESIII Collaboration),arXiv:1107.1806.

\end{thebibliography}
\end{document}